\documentclass[manuscript]{aastex}

\usepackage{epsfig}
\usepackage{bm}
\usepackage{subfigure}
\usepackage{multicol}
\usepackage{multirow}
\usepackage{array}
\newcommand{\PreserveBackslash}[1]{\let\temp=\\#1\let\\=\temp}
\newcolumntype{C}[1]{>{\PreserveBackslash\centering}p{#1}}
\newcolumntype{R}[1]{>{\PreserveBackslash\raggedleft}p{#1}}
\newcolumntype{L}[1]{>{\PreserveBackslash\raggedright}p{#1}}

\shortauthors{Song & Zhang} \shorttitle{On the relationship between flare-related structure and magnetic field changes}

\begin{document}

\title{On the relationship between sunspot structure and magnetic field changes associated with solar flares}

\author{Y. L. Song\altaffilmark{1,2} \& M. Zhang\altaffilmark{1}}

\altaffiltext{1}{Key Laboratory of Solar Activity, National Astronomical Observatories, Chinese Academy of
 Sciences, A20 Datun Road, Chaoyang District, Beijing 100012, China; \\ Email: ylsong@bao.ac.cn}
\altaffiltext{2}{University of Chinese Academy of Sciences, China}

\begin{abstract}
Many previous studies have shown that magnetic fields as well as sunspot structures present rapid and irreversible changes associated with solar flares. In this paper we first use five X-class flares observed by SDO/HMI to show that not only the magnetic fields and sunspot structures do show rapid, irreversible changes but also these changes are closely related, both spatially and temporally. The magnitudes of the correlation coefficients between the temporal variations of horizontal magnetic field and sunspot intensity are all larger than 0.90, with a maximum value of 0.99 and an average value of 0.96. Then using four active regions in quiescent times, three observed and one simulated, we show that in sunspot penumbra regions there also exists a close correlation between sunspot intensity and horizontal magnetic field strength, in addition to the well-known one between sunspot intensity and normal magnetic field strength. Connecting these two observational phenomena, we show that the sunspot structure change and the magnetic field change are the two facets of the same phenomena of solar flares, one change might be induced by the change of the other due to a linear correlation between sunspot intensity and magnetic field strength out of a local force balance.
\end{abstract}

\keywords{Sun: activity --- Sun: evolution --- Sun: flares --- Sun: magnetic fields --- Sun: sunspots}


\section{Introduction}

Since Hale (1908) discovered the existence of magnetic fields in sunspots, a consensus has been reached that the magnetic fields play an important role in various forms of solar activities such as flares (Wang \& Liu 2015). It is found that there is a close correlation between the complexity of magnetic fields and the productivity of eruptive events (Zirin \& Liggett 1987; Wang et al. 1991; Song et al. 2013). The evolution of magnetic field topological structure is believed to result in the transport of magnetic energy from the solar interior to the surface and then the storage of magnetic energy in the corona (Berger 1988; Zhang \& Low 2005).

Many early studies (Severny 1964; Zvereva \& Severny 1970; Tanaka 1978; Patterson 1984; Chen et al. 1989) have already found that there are changes of magnetic fields associated with solar flares. These changes are classified into two categories: one is the so-called magnetic transient (e.g. Kosovichev \& Zharkova 2001; Qiu \& Gary 2003; Zhao et al. 2009) and the other is the rapid and permanent change (e.g. Wang 1992; Wang et al. 1994; Kosovichev \& Zharkova 2001; Wang et al.2002a, 2002b, 2004). Magnetic transients are found in regions near the flare footpoints and they usually persist for a very short time. A general belief is that the flare-associated magnetic transients are resulted from the observational effect produced by the changes of spectral profiles. For the flare-associated rapid and permanent changes of magnetic fields, they are considered as real changes of the magnetic fields.

Many researchers have made in-depth studies on the flare-associated permanent changes of magnetic fields (e.g. Cameron \& Sammis 1999; Spirock et al. 2002; Wang et al. 2002b; Wang 2006). Using GONG (Global Oscillation Network Group, http://gong.nso.edu/; Harvey et al. 1988) data, Sudol \& Harvey (2005) investigated the changes of longitudinal magnetic fields in 15 X-class flares and found that there exist stepwise and persistent changes of magnetic fields in the areas that are associated with flares. Using line-of-sight (LOS) magnetic field data, especially SOHO/MDI (Scherrer et al. 1995), many researchers have found that the magnetic fluxes usually increase in the flaring regions towards the limb but decrease in regions towards the disk center (Yurchyshyn et al. 2004; Wang 2006; Wang \& Liu 2010). This phenomenon suggests that the overall magnetic fields of active regions may become more flatter after flares (Wang \& Liu 2010;  Wang \& Liu 2015). Using high temporal and spatial resolution SDO/HMI (Scherrer et al. 2012;  Schou et al. 2012a, 2012b) vector magnetic field data, Liu et. al (2012) reported a rapid and permanent change of magnetic field during the M6.6 flare of NOAA 11158 on 2011 February 13. They found that the mean strength of horizontal magnetic field has increased about 28\% in the flaring polarity-inversion-line (PIL) region. Similarly, Wang et. al (2012) studied 18 flares in four active regions and have found similar changes of magnetic field around the flaring PILs.

Flare-associated changes in white-light continuum intensities have also been found in previous studies (Wang et al.2004; Deng et al. 2005; Liu et al. 2005; Chen et al. 2007; Li et al. 2009; Su et al. 2011). It is often found that regions in the outer penumbra of a $\delta$-spot active region become brighter after the flare whereas regions in the inner penumbra that is near to the flaring PIL become darker after the flare (e.g. Deng et al. 2005; Li et al. 2009; Wang et al. 2012; Wang et al. 2013). For example, Liu et al. (2005) analyzed six X-class and one M-class flares to study the changes happening/occurring in the sunspot structure during the flare. They demonstrated that these flare-associated changes were irreversible, and suggested that the reason for the decay or enhancement of the penumbrae intensity was related to the magnetic fields in penumbral regions become more vertical or horizontal.

The objective of our work here is to further study the changes of magnetic fields and the changes of sunspot (intensity) structures associated with solar flares, with a particular interest in studying and understanding the relationship between the two changes. In Section 2 we first verify these flare-associated permanent changes as found in previous studies and study further the relationship of their temporal variations. Then in Section 3 we try to understand the close relation between the two changes by pointing out the existence of a linear relationship between the sunspot intensity and the horizontal magnetic field strength in the outer penumbra during quiescent times. A brief summary and discussion are given in Section 4.


\section{Spatial and temporal correlation between sunspot intensity change and horizontal magnetic field change}

The data we use in this study were obtained by HMI (\emph{Helioseismic and Magnetic Imager}) on board SDO (\emph{Solar Dynamics Observatory}). The HMI instrument is designed to study the oscillations and the magnetic field at the solar surface. It observes the full solar disk using filtergrams taken in six spectral positions around the line Fe I 6173 \AA, reaching a spatial resolution of $1''$ with a $4096\times4096$ CCD detector. We use the vector magnetic field ($B_x$, $B_y$ and $B_z$) and the continuum intensity ($I$) data from the series of hmi.sharp\_cea\_720s, processed by the HMI team using a cylindrical equal area (CEA) projection. The time cadence of these series of data is 12 minutes.

Note that HMI uses International Atomic Time (TAI) instead of Coordinated Universal Time (UTC). The difference between TAI and UTC is that UTC have been added leap seconds over the years due to the gradual slowing down of the Earth's rotation rate, whereas such leap seconds have not been added to TAI. For instance, from 2012 July 1 to 2015 June 30, UTC lags TAI by 35 seconds. We have changed the TAI times that HMI uses to the UTC times by subtracting those extra seconds in corresponding years in order to coordinate GOES X-ray flux observations that use UTC. Throughout this paper we use the term `UT' to represent Coordinated Universal Time (UTC).

To study the relationship between the change in the sunspot structure and its magnetic field change during the flare, we have chosen five active regions showing a complex topology. These active regions all produce several major eruptive events during their passages across the disk. They are selected based on the following two criteria. The first is that it produces a major flare (class $\ge X1.0$) and the second is that it is not too close to the solar limb when the flare takes place. Five X-class flares from these five active regions are investigated. Their information is shown in Table 1.

\begin{table}[htbp]
\centering
\centerline{\footnotesize \textbf{Table 1.} Event list}
\label{tab1}
{\scriptsize
\begin{tabular}{cccccccc}
\hline
\hline
NOAA  & \multicolumn{6}{c}{Flare} & CME \\
\cline{2-7}
No. & Date & Start & Peak & End & Class & Position &\\
\hline
11158  & 2011.02.15 & 01:44 & 01:56 & 02:06 & X2.2 & S20W10 & Y\\
11283  & 2011.09.06 & 22:12 & 22:20 & 22:24 & X2.1 & N14W18 & Y\\
11429  & 2012.03.07 & 00:02 & 00:24 & 00:40 & X5.4 & N18E26 & Y\\
12205  & 2014.11.07 & 16:53 & 17:25 & 17:34 & X1.6 & N14E36 & Y\\
12242  & 2014.12.20 & 00:11 & 00:28 & 00:55 & X1.8 & S19W29 & Y\\
\hline
\end{tabular}
}
\end{table}

Figure 1 presents the study of the X2.2 flare in active region NOAA 11158. GOES observation (Grubb 1975; Neupert 2011) shows that the flare starts at 01:44 UT on 2011 February 15, reaches the peak flux at 01:56 UT and stops at 02:06 UT. For this flare and the other four listed in Table 1, we have studied the evolution of the horizontal magnetic field and the intensity structure in corresponding active regions for 5 hours: 150 minutes before the peak time of the flare, and 150 minutes after.  Panel (a) shows the continuum intensity map of NOAA 11158, obtained by SDO/HMI at 01:23:20 UT, i.e., 36 minutes before the peak time. Panel (b) shows the normal magnetic field ($B_z$) map obtained at the same time.

Panel (c) shows the difference between the intensity image just after the flare ends (02:11 UT) and the one before the flare starts (01:23 UT). It can be seen here that there are obviously two regions with significant changes of continuum intensity: the dark Area 1 (outlined by a green rectangle) where the continuum intensity becomes smaller after the flare, and the bright Area 2 (outlined by a yellow rectangle) where the continuum intensity becomes larger after the flare. The same regions (Areas 1 and 2) have also been outlined in Panel (d) by a green and a yellow rectangles respectively.

In a similar way, Panel (d) shows the difference between the horizontal magnetic field ($B_h$) map just after the flare ends (02:11 UT) and the one before the flare starts (01:23 UT). The horizontal magnetic field maps are obtained by $B_h=\sqrt{B_x^2+B_y^2}$, showing the strength of the horizonal magnetic field. We see here that there are also dark and bright regions, `dark' means that the horizontal magnetic field strength $B_h$ becomes smaller after the flare and `bright' means that $B_h$ becomes larger after the flare. We use blue contours to outline where $\triangle {B_h}<-100 ~G$ and red contours where $\triangle {B_h}>100 ~G$. Again, the same contours are also overlaid in Panel (c).

Comparing Panel (c) and Panel (d), we see that where the continuum intensity becomes smaller after the flare (green rectangle Area 1 in Panel c) the horizontal magnetic field strength becomes larger (green rectangle Area 1 in Panel d) and where the continuum intensity becomes larger after the flare (yellow rectangle Area 2 in Panel c) the horizontal magnetic field strength becomes smaller (yellow rectangle Area 2 in Panel d). Similarly, where the horizontal magnetic field strength becomes smaller (within blue contours of $\triangle {B_h}=-100 ~G$) the continuum intensity becomes larger and where the horizontal magnetic field strength becomes larger (within red contours of $\triangle {B_h}=100 ~G$) the continuum intensity becomes smaller. These show that the changes of the horizontal magnetic field strength and the continuum intensity during the flare correlate spatially very well.

This coordinating change can be seen more clearly from the temporal variations of the continuum intensity and horizontal magnetic field strength, shown in Panels (e) - (h). Panel (e) shows the temporal variations of averaged continuum intensity (red line) and $B_h$ (green line) in Area 1, Panel (f) in Area 2, Panel (g) in area where $\triangle {B_h}>100 ~G$ and Panel (h) in area where $\triangle {B_h}<-100 ~G$. These profiles not only confirm the permanent changes of continuum intensity and horizontal magnetic field as discovered by many previous studies but also show that the changes of the continuum intensity and the horizontal magnetic field take place at near the same time around the flare peak time. That means the changes of the continuum intensity and the horizontal magnetic field also correspond temporally very well.

To show this temporal correspondence better, we have calculated the correlation coefficients between the continuum intensity variation (red line) and the $B_h$ variation (green line) in Panels (e) - (h) respectively. They are -0.97, -0.97, -0.99 and -0.96, respectively. We see that the magnitudes of the correlation coefficients are all very high, getting close to 1 with the smallest being 0.96. This proves that the two quantities, the continuum intensity and the horizontal magnetic field strength, do show strong corresponding changes temporally.

Similar to Figure 1, the evolutions of sunspot intensity and horizontal magnetic field and their relations for the other four active regions (NOAA 11283, 11429, 12205, 12242) are presented in Figures 2 -5 respectively. These four more examples all show the same result as in Figure 1: the evolutions of the sunspot continuum intensity and the horizontal magnetic field strength show coordinating changes both spatially and temporally.

By spatially, we mean that where the continuum intensity becomes smaller (or larger) after the flare, the horizontal magnetic field strength becomes larger (or smaller), and where the horizontal magnetic field strength becomes smaller (or larger) after the flare, the continuum intensity becomes larger (or smaller). By temporally, we mean that the changes of the continuum intensity and the horizontal magnetic field take place at near the same time, and the magnitudes of the correlation coefficients between the continuum intensity variation and the $B_h$ variation are all very large. The maximum magnitude of the correlation coefficients is as high as 0.99, and the lowest is still 0.90 with an average number of 0.96, as listed in Table 2.

\begin{table}[htbp]
\centering
\centerline{\footnotesize \textbf{Table 2.} Correlation coefficients between intensity variation and $B_h$ variation}
\label{tab2}
{\scriptsize
\begin{tabular}{cccccccc}
\hline
\hline
NOAA  & \multicolumn{2}{c}{Flare} & &\multicolumn{4}{c}{Correlation Coefficients} \\
\cline{2-3}\cline{5-8}
No. & Date & Class & & Area 1 & Area 2 & $\triangle B_h>$100 G & $\triangle B_h<$-100 G\\
\hline
11158  & 2011.02.15 & X2.2 & & -0.97 & -0.97 & -0.99 & -0.96\\
11283  & 2011.09.06 & X2.1 & & -0.98 & -0.98 & -0.96 & -0.98\\
11429  & 2012.03.07 & X5.4 & & -0.98 & -0.98 & -0.98 & -0.95\\
12205  & 2014.11.07 & X1.6 & & -0.94 & -0.97 & -0.96 & -0.90\\
12242  & 2014.12.20 & X1.8 & & -0.96 & -0.99 & -0.95 & -0.94\\
\hline
\end{tabular}
}
\end{table}

It is worthy of mentioning here that a correlation between the sunspot structure change and the magnetic field change is not the first time been found here. Previous studies (e.g. Deng et. al 2005; Liu et. al 2005; Wang et al. 2013) have already found that there are flare-associated structure changes together with the horizontal field strength changes. However, the purposes of their studies were mainly to address the phenomena of flare-associated penumbra decay (e.g. Deng et al. 2005; Liu et al. 2005) or penumbra formation (e.g. Wang et al. 2013), whereas ours is mainly to discuss the close relationship between the continuum intensity variation and the horizontal field strength variation. Our methods applied here exposed these changes more clearly, both in penumbra `decaying' region (where continuum intensity increases) and `formation' region (where continuum intensity decreases) in the same active region at the same time. In particular, we have shown how high the correlations are between these changes of sunspot intensity and the magnetic field. Such a high correlation implies a strong physical connection between these two changes and it stimulates us to pursue the work that we will present in the next section.

As a comparison, we have also studied the variations of continuum intensity and horizontal magnetic field in NOAA 12242 during a quiescent time period with no flares. From 18:30 UT to 23:30 UT on 2014 December 19 NOAA 12242 was in a relatively quiescent state with Goes X-ray flux about one order of magnitude lower than that during the flare. Same as in previous cases, we have chosen a period of 5 hours to study and the results are presented in Figure 6. Similar to Figures 1-5, Panels (a) and (b) are respectively the continuum intensity image and the normal magnetic field map of NOAA 12242 at 20:59 UT on 2014 December 19.

Also as before, Panels (c) and (d) show the difference images of continuum intensity and horizontal magnetic field between two different times (20:35 UT and 21:23 UT), with green and brown contours in Panel (c) showing regions where $\triangle I>2000~DN$ and $\triangle I<-2000~DN$ respectively; red and blue contours in Panel (d) showing areas where $\triangle B_h>100~G$ and $\triangle B_h<-100~G$ respectively. Area 1 outlined by the green box and Area 2 outlined by the yellow box are the same regions as the two in Figure 5. From Panels (c) and (d) we see that there are also variations of the continuum intensity and the horizontal magnetic field during quiescent times. However, unlike those changes during the flare, that is, the continuum intensity and $B_h$ change significantly in large areas near the PIL, these changes are randomly distributed as small patches with roughly equal number of positive (green and red contours) and negative (red and blue contours) changes.

Temporal variations of the continuum intensity and the horizontal magnetic field in Areas 1 and 2 are plotted respectively in Panels (e) and (f). We can see that the average values of continuum intensity and $B_h$ remain almost unchanged here when the active region is in a relatively quiescent time. This presents a strong contrast to those changes of continuum intensity and $B_h$ in these two areas when there is a flare, as shown in Figure 5.

Temporal variations of continuum intensity and horizontal magnetic field in regions $\triangle I>2000~DN$, $\triangle I<-2000~DN$, $\triangle B_h>100~G$ and $\triangle B_h<-100~G$ are also plotted respectively in Panels (g)-(j). It is interesting to see that, though the changes of the continuum intensity and the $B_h$ are randomly distributed as small patches within the whole active region, there is also an anti-correlation between the variations of the continuum intensity and the horizontal magnetic field. A calculation of the correlation coefficients between the two variation curves in these panels, results of which are listed in Table 3, show that the correlation coefficients are also high, although not as high as those in Table 2 for those X-class flares. This again implies that there may be a strong physical reason underlying the high correlation between the changes of continuum intensity and the horizontal magnetic field, regardless whether there is a flare or not. Understanding this high correlation is the main motivation of the next section.


\begin{table}[htbp]
\centering
\centerline{\footnotesize \textbf{Table 3.} Correlation coefficients between intensity variation and $B_h$ variation in quiescent times}
\label{tab3}
{\scriptsize
\begin{tabular}{ccccccccc}
\hline
\hline
NOAA  & \multicolumn{3}{c}{Time} & &\multicolumn{4}{c}{Correlation Coefficients} \\
\cline{2-4}\cline{6-9}
No. & Date & Start & End &  & $\triangle I>$2000 DN&$\triangle I<$-2000 DN&$\triangle B_h>$100 G & $\triangle B_h<$-100 G\\
\hline
12242  & 2014.12.19 & 18:30 & 23:30 & & -0.79 & -0.96 & -0.92 & -0.88\\
\hline
\end{tabular}
}
\end{table}


\section{The Intensity-Field Strength relation as the underlying physics}

The flare-associated changes of sunspot structures discovered by many previous studies as well as in ours are mainly characterized through white-light or other continuum intensity observations. These continuum intensities are in fact reflecting thermal information of the solar surface, and the relationship between the sunspot temperature and the magnetic field has been widely studied by both theories (Alfv\'en 1943; Maltby 1977) and observations (Kopp \& Rabin 1992; Solanki et al. 1993; Stanchfield et al.1997; Westendorp Plaza et al. 2001; Penn et al. 2003; Mathew et al. 2004). Usually an anti-correlation between the temperature and the field strength is found (e.g. Kopp \& Rabin 1992; Stanchfield et al. 1997; Mathew et al. 2004)\textbf{,} and the relationship can be nonlinear as found in Solanki et al. (1993), Penn et al. (2003) and Mathew et al. (2004).

Motivated by our finding that sunspot structures and magnetic fields change at nearly the same time during the flare events and their evolutions present strong anti-correlations, in this section, we study the relationship between the continuum intensity and the field strength, instead of the one between the temperature and the field strength. By physics the observed continuum intensity is determined by the temperature of the solar surface, but the two quantities are not linearly related. We notice that a nonlinear relationship between the temperature and the field-strength in the whole sunspot does not necessarily imply a nonlinear relation between the continuum intensity and the field-strength, particularly if considering the relation in sunspot umbra and penumbra separately. Because those flare-associated changes are always found in sunspot outer penumbra near PILs, we are motivated to study the continuum intensity - field strength relation in sunspot outer penumbra rather than in the whole sunspot.

Figure 7 presents such a study for two simple active regions, left panels for NOAA 11084 and right panels for a numerically simulated sunspot. As in previous section, NOAA 11084 is chosen as a representative of a simple active region during a quiescent time with no activities and the HMI hmi.sharp\_cea\_720s series data at a moment that the active region is closest to the disk center is used. The continuum intensity and the normal magnetic field $B_z$ maps are presented in Panels (a) and (b) respectively. The white contours in these images outline where the continuum intensity is of $3.5\times 10^4$ DN and only regions outside the contour where outer penumbra lies are studied.

Panels (c) to (e) present the relations between the normalized intensity ($I/I_0$) and the normal magnetic field ($B_z$) strength, the horizontal magnetic field strength ($B_h$) and the total magnetic field strength ($B$) respectively. By normalized intensity, the continuum intensity ($I$) in each pixel has been divided by the mean intensity of the quiet Sun $I_0$, where $I_0$ is the mean intensity of data points whose total magnetic field strengths are less than 50 G. Each dot in the scatter diagrams presents a data point in the outer penumbra region. For data points in each bin of 50 G an average of the normalized intensity is obtained and these averages form the red lines in these diagrams. The green line in each diagram presents the result of a linear fitting. We see that the normalized intensity is related to the normal field strength, the horizontal field strength and the total field strength all linearly. The slopes of the fitting lines are almost the same and the correlation coefficients are all high (-0.80, -0.94 and -0.93 for Panels (c), (d) and (e) respectively). This tells us that the linear relationship preserves very well in these outer penumbra regions, not just for the normal magnetic field strength that most previous studies focused on, but also for horizontal magnetic field strength which we are mostly interested in.

The right panels in Figure 7 present a similar study but for a numerically simulated sunspot in Rempel (2012). We choose a numerically simulated sunspot to study in order to convince ourself that the linearity we found are not dependent on a particular measurement by a particular instrument. The numerical methods in Rempel (2012) considered nearly all relevant physical processes: compressible magnetohydrodynamics, partial ionization, and radiative energy transport, as presented in detail in Rempel et al. (2009a, 2009b). We use the bolometric intensity and the vector magnetic field at $\tau = 1$. The intensity map and the normal field map are presented in Panels (f) and (g) respectively. The field of view is $49\times49 ~Mm^2$, with a spatial resolution of $96 ~km$ per pixel. Because of the low intensity contrast in the numerically simulated sunspot, we simply use a white circle as outlined to exclude the umbra regions. Same as for the case of NOAA 11084, only outer penumbra regions outside the circle are studied.

Similar to Panels (c) to (f), Panels (h) to (j) present the relationship between the normalized intensity ($I/I_0$) and the normal magnetic field ($B_z$) strength, the horizontal magnetic field strength ($B_h$) and the total magnetic field strength ($B$), respectively, for the simulated sunspot. Again we see that the linear relationship preserves very well in the outer penumbra regions, even for this simulated sunspot. The slops of the linear fitting are the same for $B_z$, $B_h$ and $B$. The correlation coefficients are also high. More impressive is that the linear fitting results are almost the same for the observed sunspot and the simulated sunspot. This consistency supports the evidence of the linearity we found in this study.

Because active regions we studied in previous section are all complicated active regions, here we choose two of them to study, to show that the linearity we found also exist in complicated active regions. Figure 8 presents a similar study as in Figure 7 but for two complicated active regions, NOAA 11158 and NOAA 11283. Different from our interest in previous section, here we choose a moment for each of these two active regions when they are closest to the disk center and have no any flare within two hours around the studied time. Here left panels present results for NOAA 11158 and right panels for NOAA 11283. As before, to exclude the umbra regions, we have used red contours in the top two panels to outline where continuum intensity is of $3.5\times 10^4$ DN. We only study the data points within the black rectangles and outside the red contours, to represent outer penumbra regions. The scatter diagrams of the data points between the normalized intensity and $B_z$, $B_h$ and $B$ are respectively plotted in Panels (c) to (e) and Panels (h) to (j), as in Figure 7. We see here that even for complicated active regions at quiescent times the same linearity presents. Again, the linear fitting results are almost the same for the observed complicated active regions, simple sunspot and the numerically simulated sunspot.

It is worthy of mentioning here that the spatial resolution of SDO/HMI data has prevented us (at least partly) to investigate the fine structures of the sunspot penumbra. Many researches (see Solanki 2003 for an excellent review) have shown that the penumbra has a darker and brighter filamentary pattern with darker structures usually having lager inclinations. These fluted magnetic field in the penumbra is often called `uncombed' magnetic field on small scales, possibly caused by near-surface small-scale magneto-convections as discussed in observations (e.g. Scharmer et al. 2011) and simulations (e.g. Rempel et al. 2009a, 2009b). The relationship between the variations of the fluted magnetic fields and their intensities traveling along a circle around the center of the umbra will be interesting to study, but it requests data with higher spatial resolutions such as those obtained by Hinode/SP. Here they possibly just contribute to the scatters in Panels (c) to (j) of Figures 7 and 8 and lead to a looser linear relationship between the intensity and the magnetic field strength.

Though these linear fittings are not perfect, but they do give a consistent fitting result between different active regions, observed or even simulated. These linear relationships imply that the sunspot intensity and the magnetic field strength are highly correlated, possibly due to a magneto-hydrostatic equilibrium in sunspots as discussed in Maltby (1977) and many others. We argue that this linear relationship may be the underlying reason why the sunspot structure and the magnetic field change so coordinately, both spatially and temporally. We argue that the sunspot structure change and the magnetic field change are the two facets of the same phenomena of solar flares, one change might be induced by the change of the other. Of course, our study here is not capable of addressing which one induces the other. But we speculate that the change of the magnetic field during the flare be induced by the magnetic impolsion as suggested by Hudson et al. (2008) and this change of the magnetic field then induce the change of the sunspot structure due to the linear relationship between the two.


\section{Summary and discussion}

This work studied the flare-associated changes of sunspot structure and magnetic field, as well as the intensity - field strength relation. Both studied fields have been extensively explored by many previous researchers, but our approach here is a slight different by trying to connect these two seemingly unrelated phenomena.

We first studied the changes of sunspot structures and magnetic fields associated with five X-class flares in five active regions. Our analysis not only confirms the results of previous studies that the magnetic fields and sunspot structures do show rapid, irreversible changes during the flares but also show in a more clear way that the changes take place coordinately both spatially and temporally. For regions showing increases (or decreases) of sunspot intensities after the flare, the horizontal magnetic fields always decrease (or increase) after the flare; for regions where horizontal magnetic fields increase (or decrease) after the flare, the sunspot intensities decrease (or increase) accordingly. Notably, the magnitudes of the correlation coefficients between the temporal variations of the horizontal magnetic field strength and the sunspot intensity are all larger than 0.90, with a maximum value of 0.99 and an average value of 0.96.

We also used four active regions at quiescent times, three observed (NOAA 11084, NOAA 11158 and NOAA 11283) and one numerically simulated, to investigate the relationship between the continuum intensity and the magnetic field strength in sunspot penumbra regions. We found that in these regions there is a close correlation between the sunspot intensity and the horizontal magnetic field strength in addition to the well-known one between the sunspot intensity and the normal magnetic field strength.

This linear relationship between the sunspot intensity and the magnetic field strength could be understood as the result of a magneto-hydrostatic equilibrium in sunspots as discussed in Maltby (1977) and others. If this relationship exists, regardless of whether in quiescent times or during a flare, then a close relationship between the continuum intensity change and the magnetic field change, with very high correlation coefficients as presented in this study, becomes quite natural and understandable.

With this relationship in mind, we speculate that we possibly do not need to find the sources of continuum intensity change and magnetic field change separately. The change of one could be induced by the change of the other, as a response to the request of a local magneto-hydrostatic equilibrium. We further speculate that we may even by-pass those difficulties, as addressed in Kleint et al. (2016), of explaining white light flares. The enhanced continuum brightness observed in the flare might be induced by a change of the magnetic field. The required energy for the white-light emission may not necessarily be transported from the overlying corona to the photosphere, but could be pulled out from a local thermal reservoir such as the convective energy on the photosphere, due to the request of a local magneto-hydrostatic equilibrium.

\acknowledgements

We are grateful to Matthias Rempel of the High Altitude Observatory for providing his numerical data we used in this study. We also thank the anonymous referee for helpful comments and suggestions that improved the paper. This work was supported by the National Natural Science Foundation of China (Grants No. 11125314, 11303052, 11303048, U1531247) and the Strategic Priority Research Program (Grant No. XDB09000000) of the Chinese Academy of Sciences, the Young Researcher Grant of the National Astronomical Observatories of CAS.


\newpage

\begin{figure}[!ht]
\centerline{\includegraphics[width=0.9\textwidth]{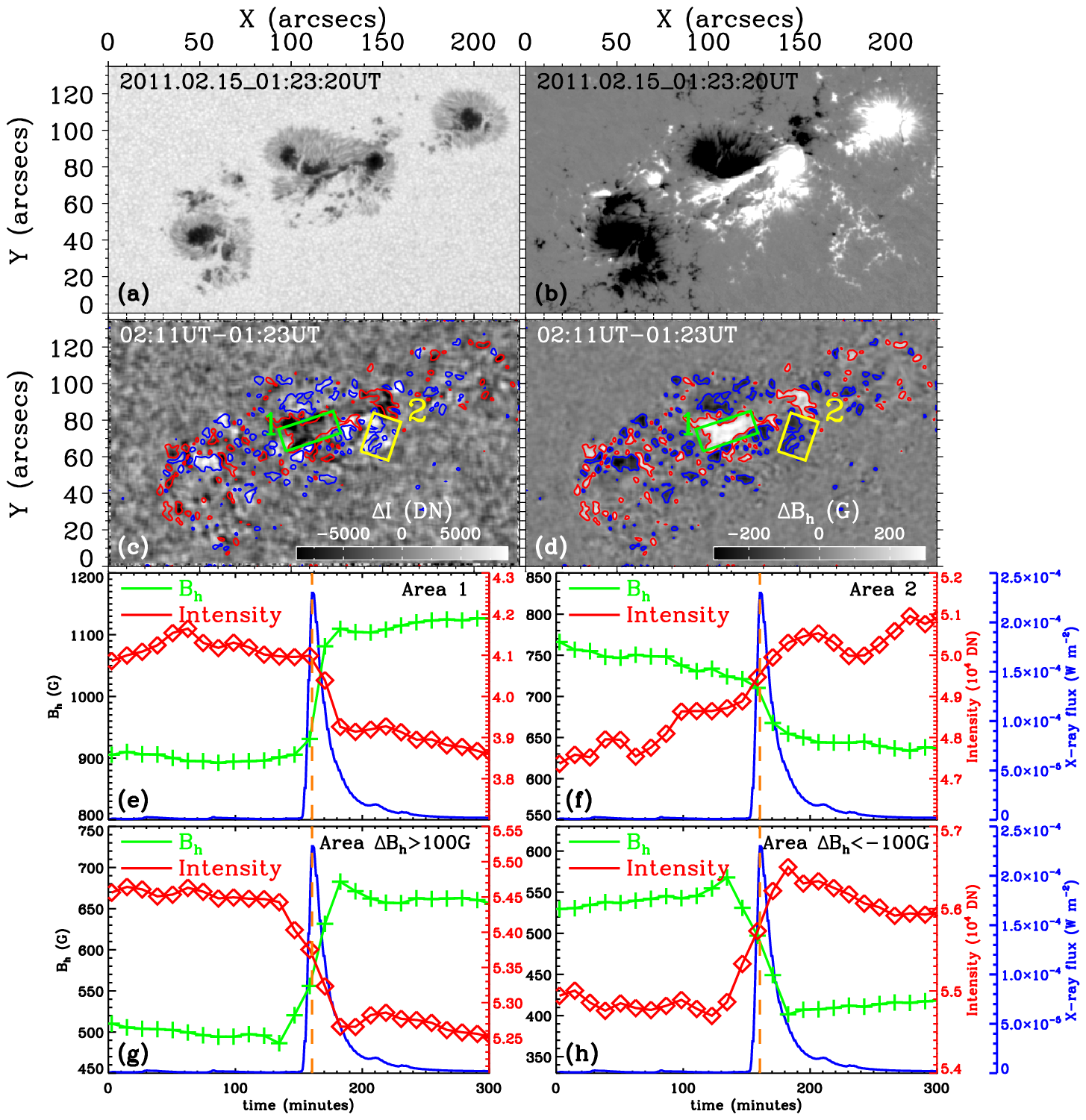}}
\caption{(\emph a): continuum intensity map of NOAA 11158; (\emph b): normal magnetic field map; (\emph c): difference image of continuum intensity; (\emph d): difference image of horizontal magnetic field; Area 1 (or 2): region where the continuum image becomes darker (or brighter) after the flare; Contours in Panels (c) and (d): $\pm$ 100 G of horizontal magnetic field change; (\emph e)-(\emph h): temporal variations of mean horizontal magnetic field strength ($B_{h}$, \emph {green} line) and continuum intensity (\emph {red} line) in Areas 1, 2, regions where $\triangle B_h>100~G$ and regions where $\triangle B_h<-100~G$, respectively. Blue solid line in Panels (e) to (h) shows the GOES X-ray flux. See text for more details.}
\end{figure}

\begin{figure}[!ht]
\centerline{\includegraphics[width=0.9\textwidth]{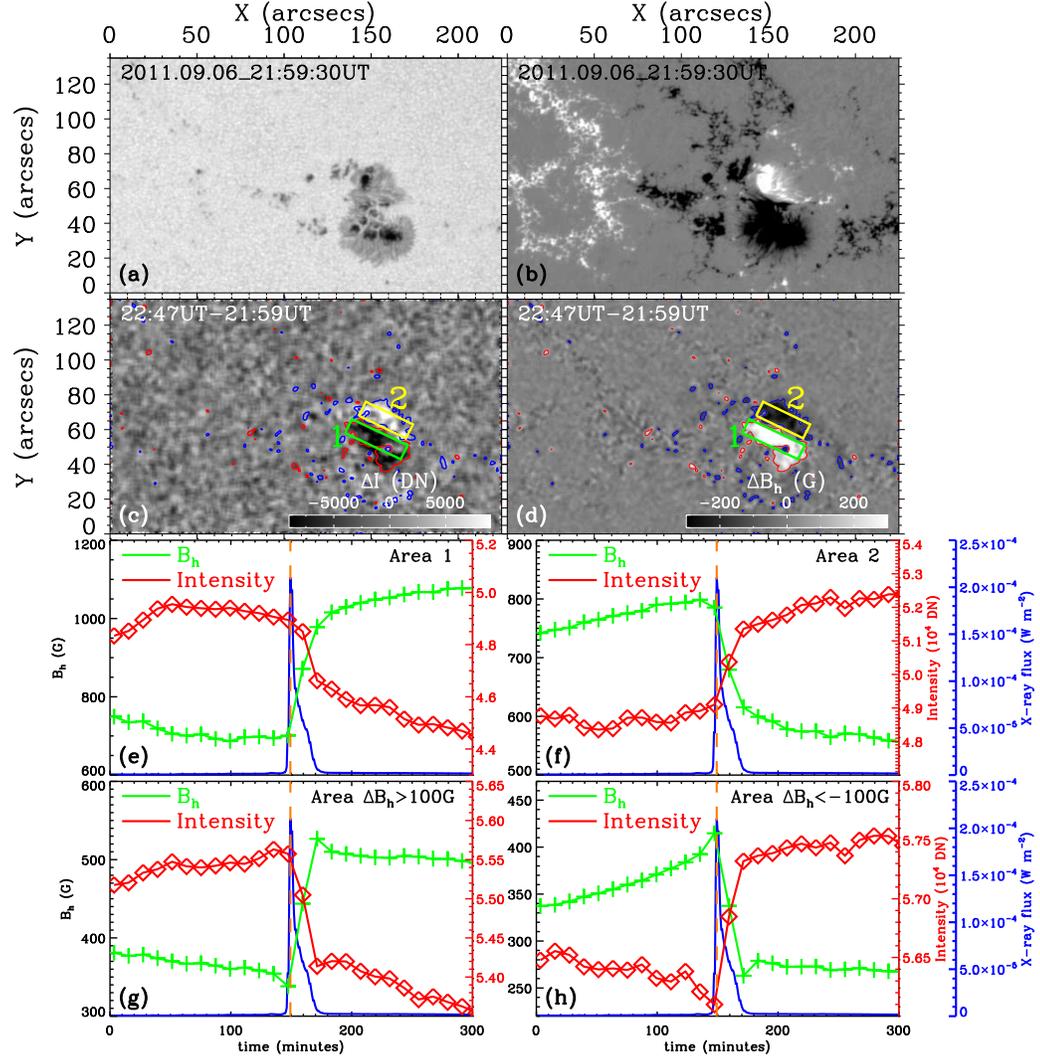}}
\caption{ Same as Figure 1 but for the X2.1 event of NOAA 11283.}
\end{figure}

\begin{figure}[!ht]
\centerline{\includegraphics[width=0.9\textwidth]{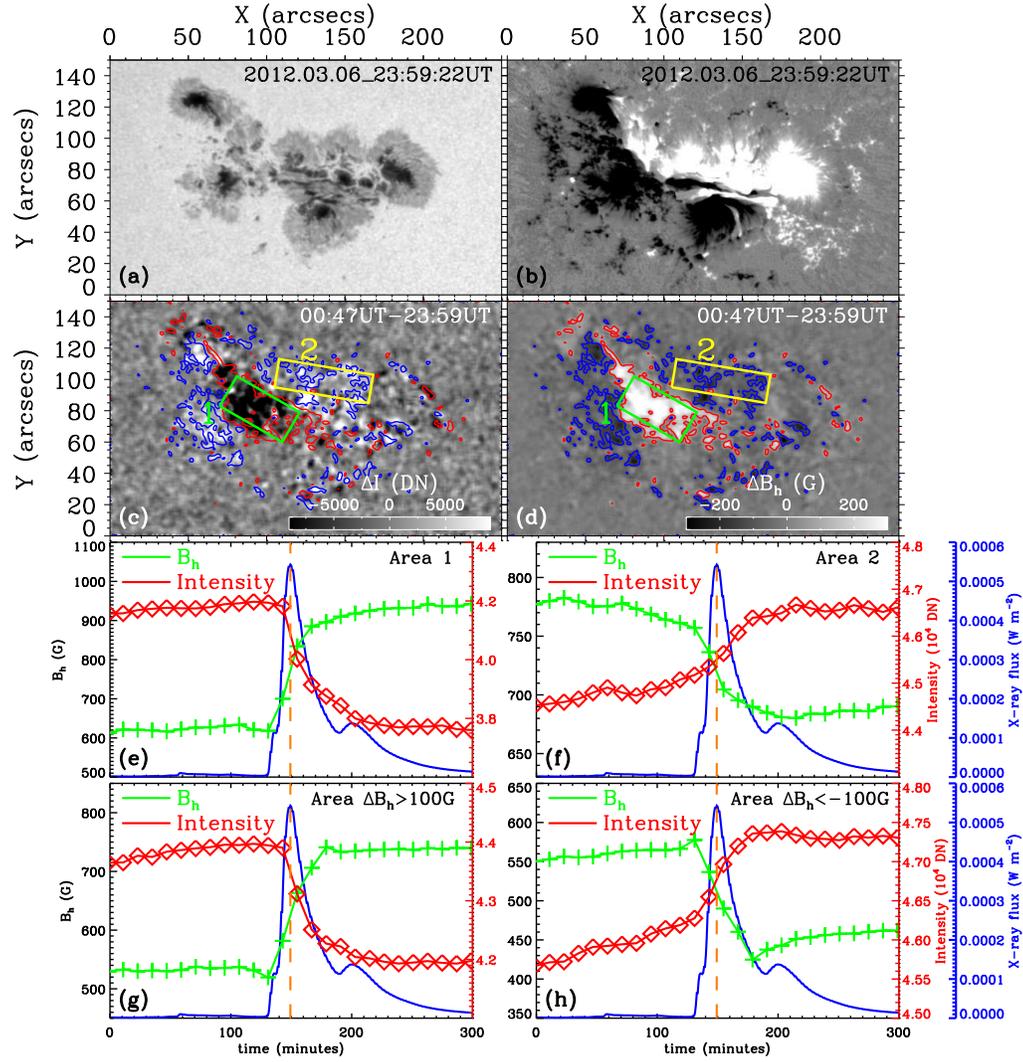}}
\caption{Same as Figure 1 but for the X5.4 event of NOAA 11429.}
\end{figure}

\begin{figure}[!ht]
\centerline{\includegraphics[width=0.9\textwidth]{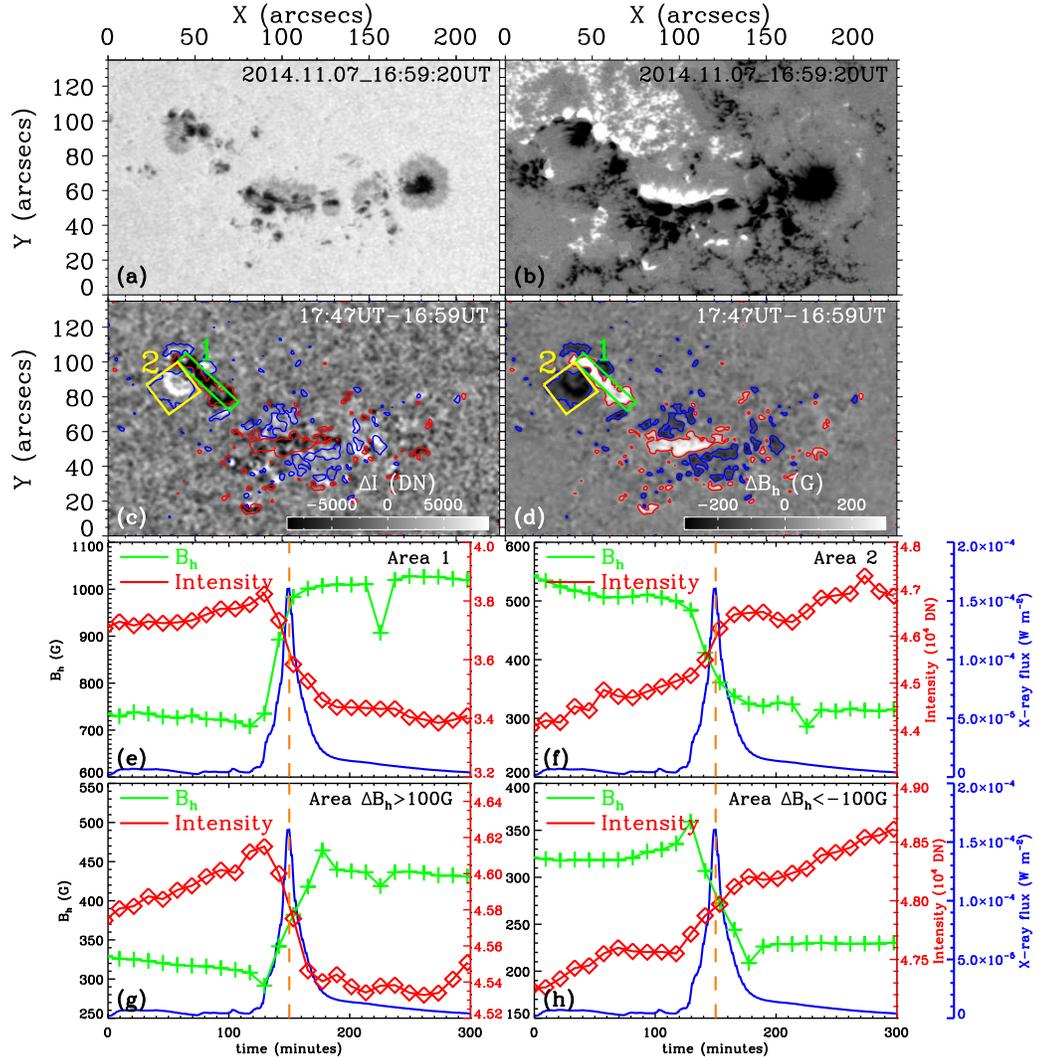}}
\caption{Same as Figure 1 but for the X1.6 event of NOAA 12205.}
\end{figure}

\begin{figure}[!ht]
\centerline{\includegraphics[width=0.9\textwidth]{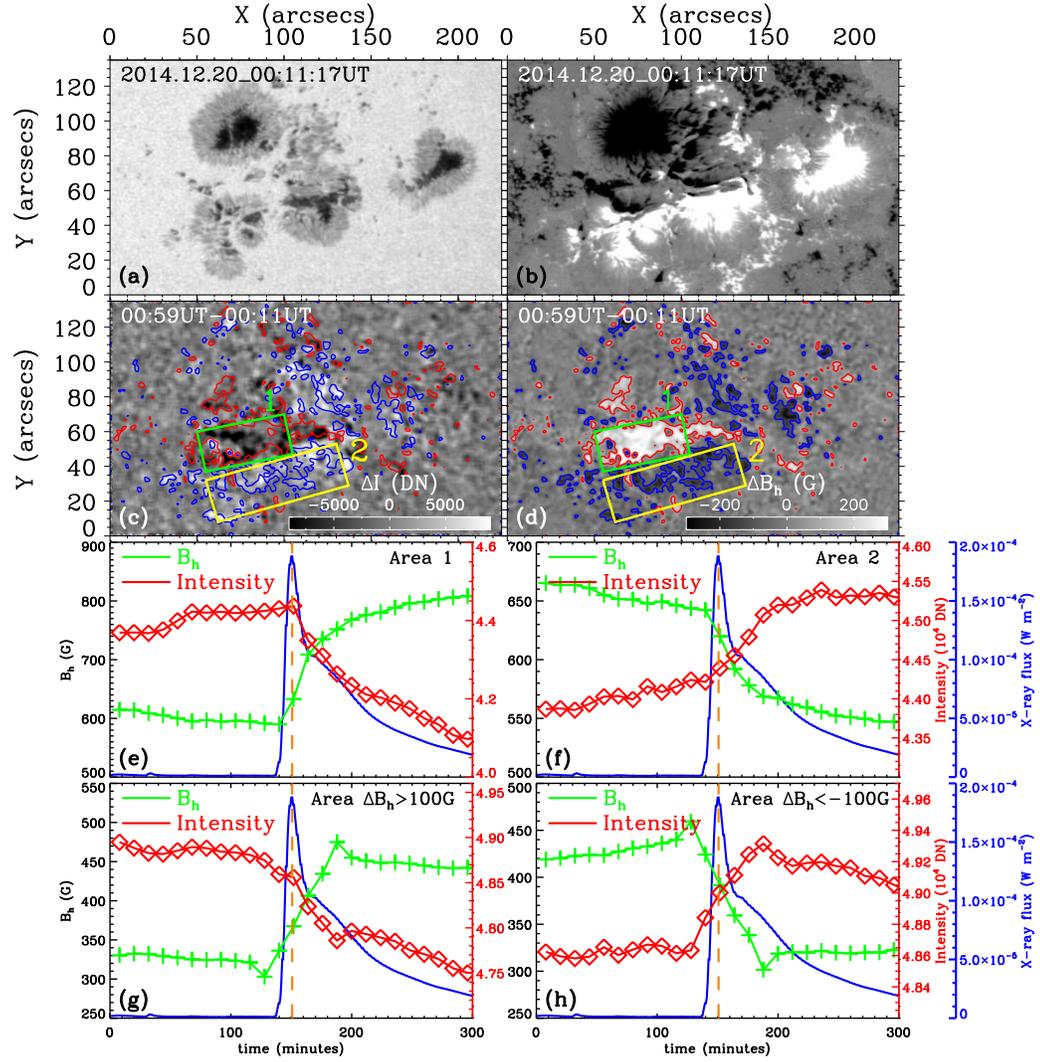}}
\caption{Same as Figure 1 but for the X1.8 event of NOAA 12242.}
\end{figure}

\begin{figure}[!ht]
\centerline{\includegraphics[width=0.8\textwidth]{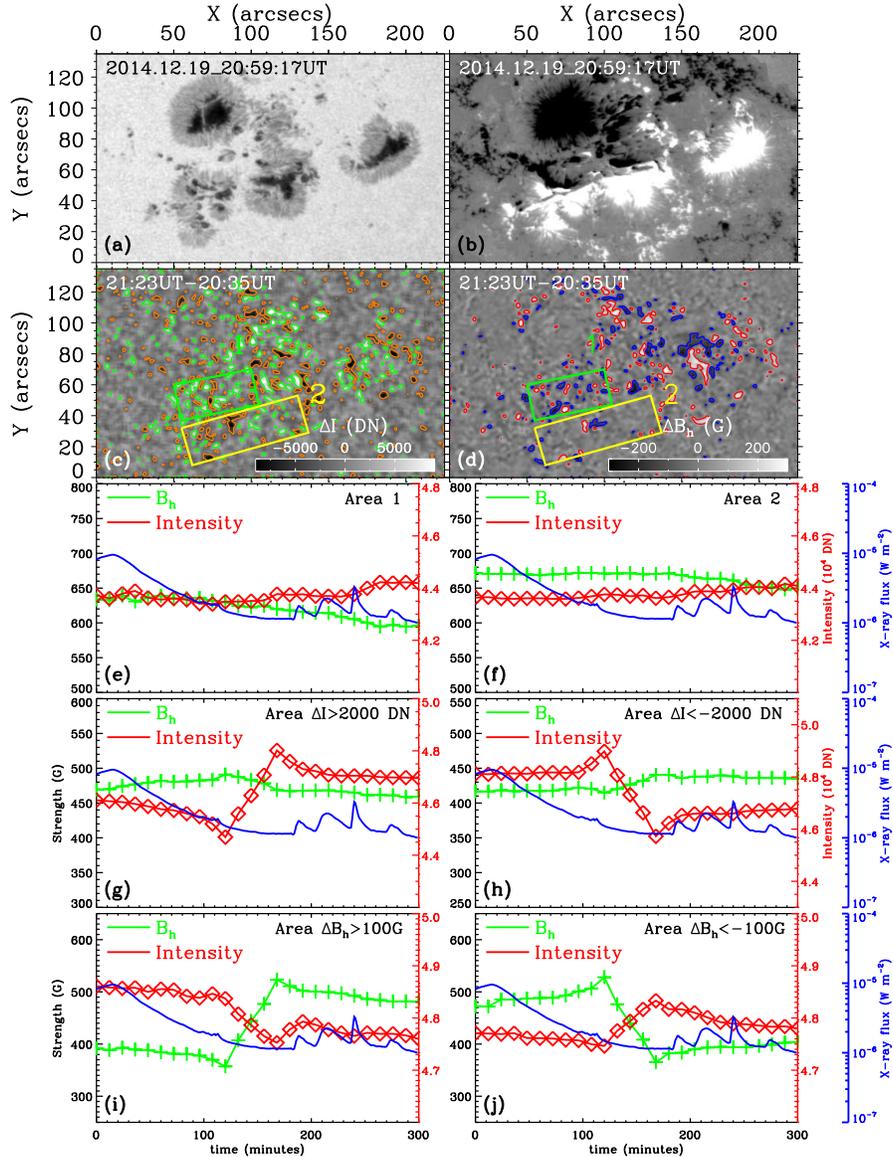}}
\caption{Similar to Figure 5 but for the NOAA 12242 during a time period with no flares. (\emph a): continuum intensity image; (\emph b): normal magnetic field image;  (\emph c): difference image of continuum intensity, contours of $\pm$ 2000 DN changes; (\emph d): difference image of horizontal magnetic field, contours of $\pm$ 100 G changes, Areas 1 and 2 same to those in Figure 5; (\emph e)-(\emph j): temporal variations of mean horizontal magnetic field strength ($B_{h}$, \emph {green} line) and continuum intensity (\emph {red} line) in regions Area 1, Area 2, where $\triangle I>2000~DN$ ,  $\triangle I<-2000~DN$ and where $\triangle B_h>100~G$ , $\triangle B_h<-100~G$, respectively. Blue solid line shows the GOES X-ray flux. The start time in (\emph e)-(\emph h) is 18:30 UT on 2014 December 19.}
\end{figure}

\begin{figure}[!ht]
\centerline{\includegraphics[width=0.9\textwidth]{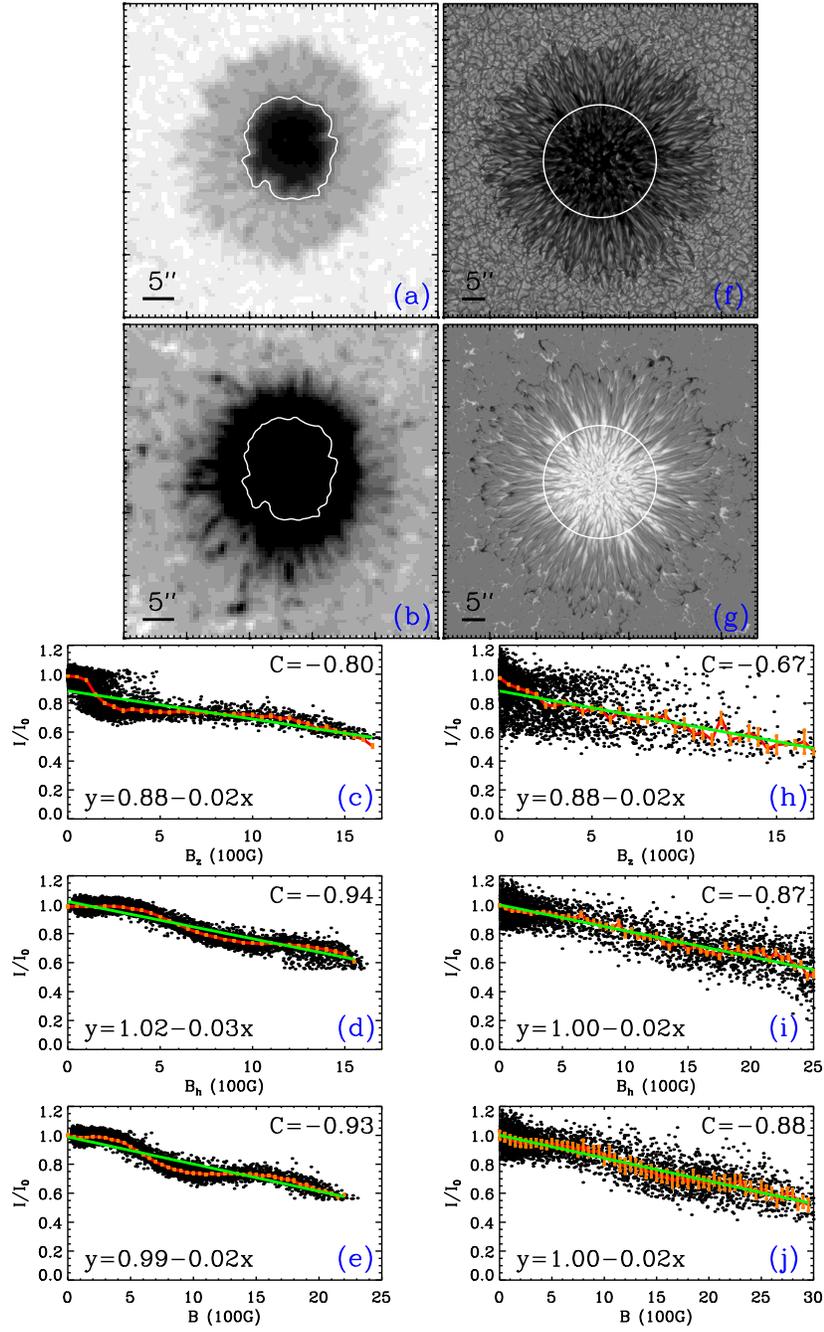}}
\caption{The relations between the continuum intensity and the magnetic field strength in outer penumbra regions (outside the contours or circles in top two panels). Left panels: of a simple sunspot in NOAA 11084; Right panels: of a numerically simulated sunspot. Panels (a) and (f): continuum intensity maps; Panels (b) and (g): $B_z$ maps. Panels (c) and (h): relations between the normal magnetic field strength and the normalized continuum intensity; Panels (d) and (i): relations between the horizontal magnetic field strength and the normalized continuum intensity; Panels (e) and (j): relations between the total magnetic field strength and the normalized continuum intensity. See text for more details.}
\end{figure}

\begin{figure}[!ht]
\centerline{\includegraphics[width=0.9\textwidth]{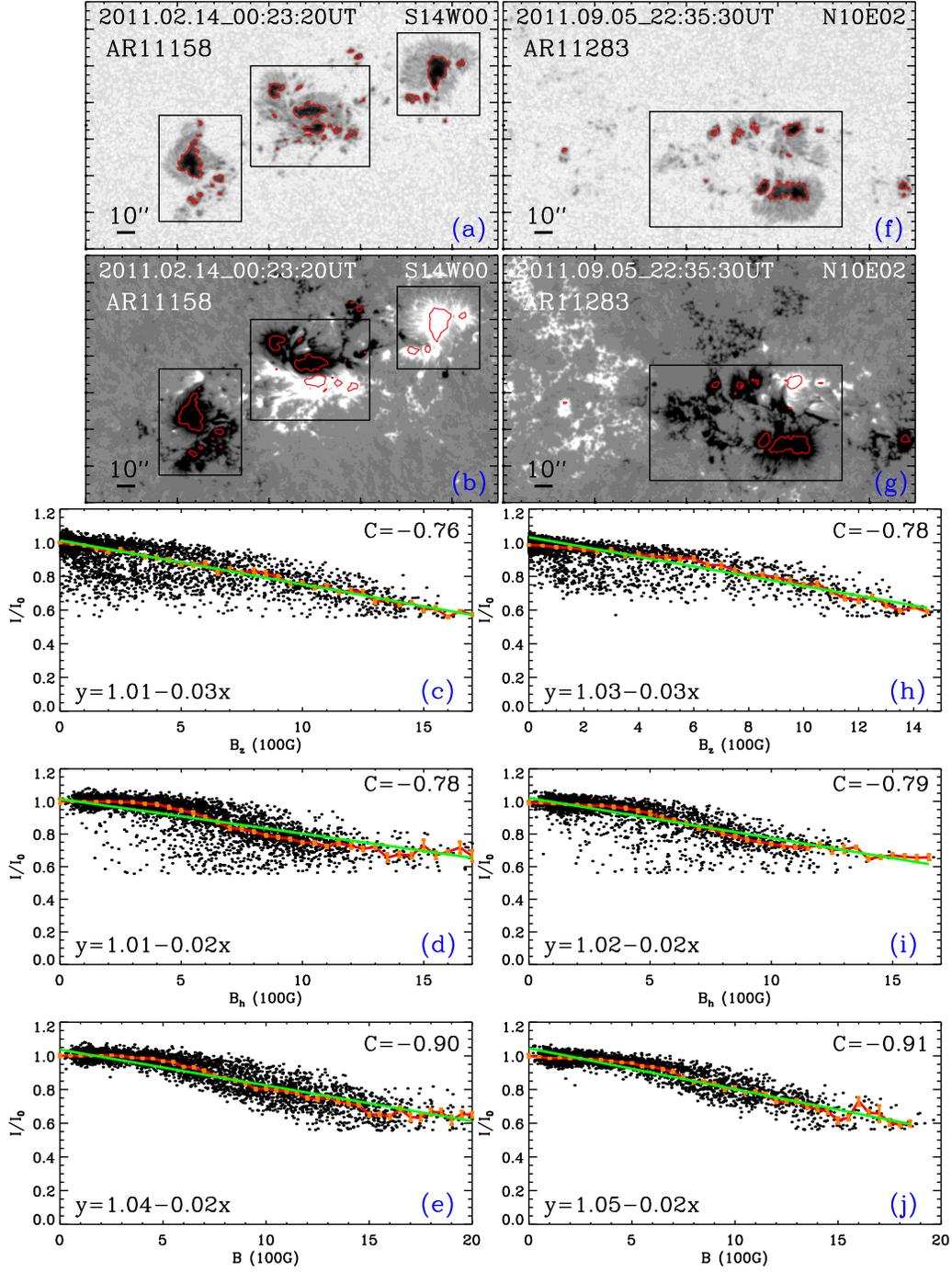}}
\caption{Similar to Figure 7 but for two complicated active regions, NOAA 11158 in left panels and NOAA 11283 in right panels.  The regions studied are outside the red contours and within the black boxes, as outlined in top two panels. See text for more details.}
\end{figure}

\end{document}